\documentclass{ifacconf}

\usepackage{graphicx}      
\usepackage{natbib}        

\usepackage{amssymb}
\usepackage{amsmath}
\usepackage{amssymb}
\usepackage{amsfonts}
\usepackage{dsfont}
\usepackage{subfigure}
\usepackage{amsbsy}
\usepackage{mathrsfs}
\usepackage{lipsum}
\usepackage{color}
\usepackage{lineno}
\usepackage{comment}

\definecolor{aoenglish}{rgb}{0.0, 0.5, 0.0}
\definecolor{darkblue}{rgb}{0.0, 0.0, 0.55}
\definecolor{darkmagenta}{rgb}{0.55, 0.0, 0.55}
\definecolor{electricviolet}{rgb}{0.56, 0.0, 1.0}
\definecolor{electricyellow}{rgb}{1.0, 1.0, 0.0}
\definecolor{forestgreen}{rgb}{0.13, 0.55, 0.13}
\definecolor{fuchsia}{rgb}{1.0, 0.0, 1.0}
\definecolor{gamboge}{rgb}{0.89, 0.61, 0.06}
\definecolor{goldenpoppy}{rgb}{0.99, 0.76, 0.0}
\definecolor{indigo}{rgb}{0.29, 0.0, 0.51}
\definecolor{internationalorange}{rgb}{1.0, 0.31, 0.0}
\definecolor{lava}{rgb}{0.81, 0.06, 0.13}
\definecolor{selectiveyellow}{rgb}{1.0, 0.73, 0.0}
\definecolor{turquoiseblue}{rgb}{0.0, 1.0, 0.94}
\definecolor{turquoise}{rgb}{0.19, 0.84, 0.78}

\newcommand{\VBr}[1]{\textcolor{black}{#1}}
\newcommand{\MFa}[1]{\textcolor{black}{#1}}
\newcommand{\SFo}[1]{\textcolor{black}{#1}}
\newcommand{\ACh}[1]{\textcolor{black}{#1}}

\def \E {\mathbf{E}}

\def \S {\mathbf{S}} 
 
\def \U {\mathbf{U}}

\def\Hmc{\mathcal{H}}

\renewcommand{\vec}[3]{\mathrm{vec}_{#1}^{#2}(#3)} 

\begin{document}
\begin{frontmatter}

\title{Uncertainty-aware data-driven predictive control in a stochastic setting\thanksref{footnoteinfo}} 

\thanks[footnoteinfo]{This project was partially supported by the Italian Ministry of University and Research under the PRIN’17 project \textquotedblleft Data-driven learning of constrained control systems\textquotedblright, contract no. 2017J89ARP.}

\author[Tue]{V. Breschi}, 
\author[UniPd]{M. Fabris}, 
\author[PoliMi]{S. Formentin},
\author[UniPd]{A. Chiuso}


\address[Tue]{Department of Electrical Engineering, 5600 MB Eindhoven, Netherlands (e-mail: v.breschi@tue.nl).}

\address[PoliMi]{Department of Electronics, Information and Bioengineering, Politecnico di Milano, via G. Ponzio 34/5, Milano, 20133, Italy (e-mail: simone.formentin@polimi.it).}
   
\address[UniPd]{Department of Information Engineering, University of Padova, via Gradenigo 6/B, Padua, 35131, Italy. (e-mail: marco.fabris.1@unipd.it, alessandro.chiuso@unipd.it)}

\begin{abstract}                
Data-Driven Predictive Control (DDPC) has been recently proposed as an effective alternative to traditional Model Predictive Control (MPC), in that the same constrained optimization problem can be addressed without the need to explicitly identify a full model of the plant. However, DDPC is built upon input/output trajectories. Therefore, the finite sample effect of stochastic data, due to, e.g., measurement noise, may have a detrimental impact on closed-loop performance. Exploiting 
a formal statistical analysis of the prediction error, in this paper we propose the first \emph{systematic approach} to deal with uncertainty due to finite sample effects. To this end, we introduce two regularization strategies \SFo{for which}, differently from existing regularization-based DDPC techniques, we propose a tuning rationale \SFo{allowing} us to select the regularization hyper-parameters \emph{before} closing the loop and \emph{without} additional experiments. \SFo{Simulation} results confirm the potential of the proposed strategy when closing the loop.
\end{abstract}


\end{frontmatter}

\section{Introduction}
Among advanced control strategies, Model Predictive Control (MPC) is nowadays one of the \SFo{most widely} employed in practice, thanks to its intrinsic ability to handle constraints, time-varying dynamics and multiple (potentially conflicting) objectives, see e.g., \cite{borrelli2017predictive}. Nonetheless, the ultimate performance attained in closed-loop with MPC critically depends on the predictive capabilities of the model featured within its optimization routine. As such, the ability of MPC to deliver the desired control performance might be jeopardized when such a mathematical description of the plant is not accurate enough. This well-known issue, that has led many research efforts towards the development of robust and adaptive MPC solutions (see, e.g., \cite{ding2017robust}), is particularly relevant when no mathematical model of the plant is available. In this case, system identification can come of help in allowing one to retrieve an accurate model of the system from data. Alternatively, in such a data-driven context, the unavoidable uncertainty of models can be dealt with by \textit{skipping} an explicit modelling step, using data to directly map the control law.  

One of the key ideas to make this shift possible is to think of past input/output records, traditionally used in system identification as training data to learn a parametric dynamical model, as a \textit{nonparametric} description of its dynamical behavior. This concept is at the core of subspace identification (see \cite{Moonen}), as well as behavioral theory in general (see \cite{willems2013introduction}), and of the (deterministic) result in \cite{willems2005note} in particular. More specifically, this last work shows that the future behaviour of a (deterministic) dynamical system can be expressed as the linear combination of a finite set of past trajectories, provided \MFa{that} the input satisfy certain persistency of excitation conditions. This deterministic result has paved the way for the recent developments of data-driven predictive control (DDPC), see, e.g., \cite{coulson2019data} and \cite{berberich2020data}. This alternative predictive approach is proven to be equivalent to traditional MPC, if data are collected in a deterministic (noiseless) setting (see \cite{krishnan2021direct}) and, in special cases, to Subspace Predictive Control, see \cite{FAVOREEL1999,Felix2021,breschi2022role}, while its performance rapidly deteriorates as stochastic data (e.g., noisy data) are used. To make DDPC less sensitive to noise in the data, different forms of regularization have then been embedded within the DDPC scheme, and they have been proven effective in handling noise (see \cite{dorfler2022bridging} for an overview of possible regularization strategies). Nonetheless, adding regularization 
\VBr{terms} to the predictive control cost implies that suitable 
\VBr{regularization weights}
must be selected \textit{a-priori}, with a non-negligible impact on the final closed-loop performance. It follows that a proper tuning of such penalties could be performed only by means of a subsequent validation phase, which must be \VBr{carried out} 
in closed-loop. These experiments might be unsafe (and often unfeasible) for the plant, as one may even end up de-stabilizing the closed-loop.

Instead of looking at the DDPC design problem from a behavioral perspective, in this paper we look at input/output trajectories from a subspace identification oriented perspective. In particular, building upon the so-called $\gamma$-DDPC formulation presented in \cite{breschi2022role}, we propose a \emph{systematic framework} to deal with uncertainty in designing data-driven predictive controllers within a stochastic setting. Specifically, by relying on the statistical analysis of the uncertainty in the data-driven predictions, we introduce two regularization schemes to limit mismatches between the true outputs and their prediction and, ultimately, improve closed-loop performance. Differently from existing regularized DDPC 
\VBr{approaches}
, we additionally propose strategies to tune regularization parameters \emph{without} requiring closed-loop experiments. The validity of \VBr{these procedures} 
for a proper tuning of the overall scheme is shown on a benchmark simulation example.

The remainder of the paper is structured as follows. In Section~\ref{sec:background} we initially provide a summary of the main features needed to construct the $\gamma$-DDPC scheme proposed in \cite{breschi2022role}. Section~\ref{sec:problem} is then devoted to the formalization of the \SFo{problem}, i.e., the design of uncertainty-aware regularization ingredients for $\gamma$-DDPC. The statistical analysis of the data-driven multi-step predictor employed in the considered predictive scheme is provided in Section~\ref{sec:uncertainty}. In light of these results, in Section~\ref{sec:reg} we propose two alternative tuning policies for the regularization penalties. Their effectiveness is illustrated through \SFo{a numerical} case study in Section \ref{sec:examples}. The paper is ended by some concluding \SFo{remarks}.

\paragraph*{Notation.} Given a signal $w(k) \in \mathbb{R}^s$, the associated (block) Hankel matrix $W_{[t_0,t_1],N} \in \mathbb{R}^{s(t_1-t_0+1) \times N}$ is defined as:
\begin{equation}\label{eq:Hankel}
W_{[t_0,t_1],N}\!:=\!\!\frac{1}{\sqrt{N}}\!\begin{bmatrix}
w(t_0) & w(t_0\!+\!1) & \cdots & w(t_0\!+\!N\!-\!1)\\
w(t_0\!+\!1) & w(t_0\!+\!2) & \cdots & w(t_0\!+\!N)\\
\vdots & \vdots & \ddots & \vdots\\
w(t_1) &w(t_1\!+\!1) & \dots & w(t_1\!+\!N\!-\!1)  
\end{bmatrix}\!\!,
\end{equation}
while we use the shorthand $W_{t_0}:= W_{[t_0,t_0],N}$ to denote a single (block) row Hankel, namely:
\begin{equation}\label{eq:Hankel:onerow}
W_{t_0}:= \frac{1}{\sqrt{N}}\begin{bmatrix}
w(t_0) & w(t_0\!+\!1) & \cdots & w(t_0\!+\!N\!-\!1)
\end{bmatrix}. 
\end{equation}

\section{Background}\label{sec:background}
Consider an \emph{unknown} discrete-time, \emph{linear time-invariant} (LTI) stochastic plant $\mathcal{S}$. Without loss of generality, let  $\mathcal{S}$ be described in the so-called minimal (i.e., reachable and observable) \emph{innovation form}, namely
\begin{equation}\label{eq:stoc_sys}
	\begin{cases}
		x(t+1)=Ax(t)+Bu(t)+Ke(t)\\
		y(t)=Cx(t)+Du(t)+e(t), 	\end{cases} \quad  t \in \mathbb{Z}
\end{equation}
where $x(t)\in \mathbb{R}^{n}$, $u(t) \in \mathbb{R}^{m}$ and $e(t) \in \mathbb{R}^{p}$ are the state, input and  innovation process  respectively, while $y(t) \in \mathbb{R}^{p}$ is the corresponding output signal. 

Let us introduce the joint input/output process $z(t)$, given by
\begin{equation}\label{eq:z}
	z(t):=\begin{bmatrix}
		u(t)\\
		y(t)
	\end{bmatrix}\VBr{.}
\end{equation}
\VBr{Given a set of $N_{data}$ input/output pairs and, thus, the sequence $\{z(j)\}_{j=1}^{N_{data}}$,} 
let the associated Hankel matrix be
\begin{equation}\label{eq:past}
	Z_P\!\!:=\!Z_{[0,\rho-1],N},
\end{equation}
where $N:=N_{data}-T-\rho+1$, $T$ is the \textquotedblleft future horizon\textquotedblright, i.e., the \VBr{prediction} horizon when solving a predictive control problem, and $\rho$ is the \textquotedblleft past horizon\textquotedblright, shaping the number of past input/output samples used to reconstruct the state at time $t$\VBr{, when it is not directly measurable}. In addition, let us define the following input and output Hankel matrices:  
\begin{align}\label{eq:future}
	U_F\!:=&U_{[\rho,\rho+T-1],N},~Y_F\!:=\! Y_{[\rho,\rho+T\!-\!1],N}.
\end{align}
Based on \eqref{eq:stoc_sys}, the Hankel of future outputs $Y_F$ can be written as a function of the previous matrices in the form
\begin{subequations}
\begin{equation}
    {Y}_{F}=\Gamma X_{\rho} +\mathcal{H}_{d}U_{F}+ \mathcal{H}_{s}E_F, 
\end{equation}
where $E_F$ is the Hankel of future innovations, $\Gamma \in \mathbb{R}^{pT\times n}$ is the extended observability matrix associated with the system, i.e., 
\begin{equation}\label{eq:observability}
	\Gamma=\begin{bmatrix} C \\ CA \\ CA^{2}\\ \vdots \\ CA^{T-1} \end{bmatrix},
\end{equation}   
while $\mathcal{H}_{d} \in \mathbb{R}^{pT \times mT}$ and $\mathcal{H}_s \in \mathbb{R}^{pT \times pT}$ are the Toeplitz matrices formed with the Markov parameters of the system, namely
	\begin{align}
		& 	\mathcal{H}_{d} = \begin{bmatrix} 
			D & 0  & 0 & \dots & 0  \\
			CB & D  &  0 &\dots & 0 \\
			CAB & CB & D  &  \dots & 0 \\
			\vdots & \vdots  & \vdots &  \ddots & \vdots & \\
			CA^{T-2}B & CA^{T-3}B  & CA^{T-4}B & \ldots &D 
		\end{bmatrix},\\
		&\mathcal{H}_s = \begin{bmatrix} 
			I & 0  & 0 & \dots & 0 \\
			CK & I  &  0 &\dots & 0 \\
			CAK & CK & I  &  \dots & 0 \\
			\vdots & \vdots  & \vdots &  \ddots & \vdots  \\
			CA^{T-2}K & CA^{T-3}K  & CA^{T-4}K & \ldots &I 
		\end{bmatrix}. 
	\end{align}
\end{subequations}
Let us additionally define $\hat Y_F $ as the orthogonal projection of $Y_F$ onto the row space of $Z_P$ and $U_F$, i.e.,
\begin{align}
     \hat{Y}_{F}&=\Gamma \hat X_{\rho} +\mathcal{H}_{d}U_{F}+ \underbrace{\mathcal{H}_{s}\Pi_{Z_{P},U_{F}}(E_F)}_{O_P(1/\sqrt{N})} \label{eq:Proj:F}
\end{align}
where the last term vanishes\footnote{For a more formal statement on this, we refer the reader to standard literature on subspace identification.} (in probability) as $1/\sqrt{N}$.

When the matrices $(A,B,C,D,K)$ are \emph{unknown}, future outputs can still be predicted from the Hankel matrices in \eqref{eq:past}-\eqref{eq:future}. \MFa{Indeed, defining the future input and output vectors
\begin{equation}\label{eq:zinit}
\begin{matrix}
    u_{f}:=\begin{bmatrix}
    u(t)\\
    u(t+1)\\
    \vdots\\
    u(t+T-1)
    \end{bmatrix},
	\quad & y_{f}:=\begin{bmatrix}
		y(t)\\
		y(t+1)\\
		\vdots\\
		y(t+T-1)
	\end{bmatrix},
    \end{matrix}
\end{equation}
$Z_{P}, U_{F}$ and $Y_{F}$ can be used in a deterministic setting to predict, given past data and future inputs, the future outputs $y_f$ 
as follows:
\begin{equation}\label{eq:DDPC_standard}
\begin{bmatrix}
    z_{init}\\u_{f}\\ y_{f}			
    \end{bmatrix}=\begin{bmatrix}
    Z_{P}\\U_{F}\\ Y_{F}
    \end{bmatrix}\alpha,~~ \text{ where } z_{init}:=\begin{bmatrix}
    z(t-\rho)\\
    \vdots\\
    z(t-2)\\
    z(t-1)
\end{bmatrix}, 
\end{equation}}with $\alpha \in \mathbb{R}^{N}$ being the variable typically optimized in DDPC problems stemming from \cite{willems2005note}, see e.g., \cite{berberich2020data} and \cite{coulson2019data}. 

Following the same rationale of \cite{breschi2022role}, we instead reformulate the previous relationship through the LQ decomposition  of the joint input-output block Hankel matrix:
	\begin{equation}\label{eq:LQ2}
	\begin{bmatrix}
	Z_{P}\\
	U_{F}\\
	Y_{F}
	\end{bmatrix}=\begin{bmatrix}
	L_{11} & 0 & 0  \\
	L_{21} & L_{22} &  0\\
	L_{31} & L_{32} & L_{33} 
	\end{bmatrix}\begin{bmatrix}
	Q_{1}\\
	Q_{2}\\
	Q_{3}
	\end{bmatrix},
	\end{equation}
where the matrices $\{L_{ii}\}_{i=1}^{3}$ are all non-singular and $Q_{i}$ have orthonormal rows, i.e., $Q_{i}Q_{i}^{\top}=I$, for $i=1,\ldots,3$, $Q_i Q_j^\top = 0$, $i\neq j$. Combining \eqref{eq:DDPC_standard} with \eqref{eq:LQ2}, we can further retrieve the following relationship:
	\begin{equation}\label{eq:LQ}
        \begin{bmatrix}z_{init}\\u_{f}\\ y_{f}			
    \end{bmatrix}=\begin{bmatrix}
    Z_{P}\\U_{F}\\ Y_{F}
    \end{bmatrix}\alpha=
	\begin{bmatrix}
	L_{11} & 0 & 0  \\
	L_{21} & L_{22} &  0\\
	L_{31} & L_{32} & L_{33} 
	\end{bmatrix}
        \MFa{\underbrace{\begin{bmatrix}
	Q_{1}\\
	Q_{2}\\
	Q_{3}
	\end{bmatrix}\alpha}_{\gamma}}.
	\end{equation}
 This \VBr{relation} 
 allows us to establish a connection between the standard optimization variable of DDPC strategies $\alpha$, and the new parameters
\begin{equation}\label{eq:gamma}
\gamma=\begin{bmatrix}
\gamma^\top_{1} &
\gamma^\top_{2} &
\gamma^\top_{3}
\end{bmatrix}^\top,
\end{equation}
which is the starting point for the derivation of the $\gamma$-DDPC scheme proposed in \cite{breschi2022role}, and \VBr{that is }at the core of this work.
\section{Problem setting}\label{sec:problem}
Consider now the predictive control problem designed for the outputs of the system to track a given reference $y_{r}(t)$, while satisfying the constraints encoded into the sets $ \mathcal{U}$, $ \mathcal{Y}$. This control problem can be cast as 
\begin{subequations}\label{eq:RHPC_prob}
	\begin{align}
		&\underset{\{u(k)\}_{t}^{t+T-1}}{\mbox{minimize}}~\frac{1}{2}\sum_{k=t}^{t+T-1} \!\!\!\|\hat y(k)\!-\! y_{r}(k)
		\|_{Q}^{2}\!+\!\|u(k)\|_{R}^{2} \label{eq:cost}\\
		& \mbox{s.t. } \hat x(k\!+\!1)\!=\!\!A\hat x(k)\!+\!Bu(k),~k \!\in\! [t,t\!+\!T),\\
		& \qquad  \hat y(k)\!=\!C\hat x(k)+Du(k),~k \in [t,t+T),\\ 		
		&\qquad \hat x(t) = x_{init},\\ 
		&\qquad u(k) \in \mathcal{U},~\hat y(k) \in \mathcal{Y},~k \in [t,t+T),
	\end{align}
\end{subequations}
where {{$k \in \mathbb{Z}$}}, $x_{init}$ is the state at time $t$, $\hat x$ and $\hat y$ are the \textit{model-based estimates} of the deterministic components of the states and outputs, while the penalties $Q \in \mathbb{R}^{p \times p}$ and $R \in \mathbb{R}^{m \times m}$, with $Q \succeq 0$ and $R \succ 0$, are selected to trade-off between tracking performance and control effort. 


Let us now assume that the system matrices $(A,B,C,D,K)$ are \emph{unknown}, while we have access to a sequence of input/output data $\mathcal{D}_{N_{data}}=\{u(j),y(j)\}_{j=1}^{N_{data}}$. Within this context, a data-driven predictive controller with the same objectives and constraints of \eqref{eq:RHPC_prob} can be formulated as follows
\begin{subequations}\label{eq:RHPC_prob_dd_gamma}
		\begin{align}
			&\underset{\gamma_2,\gamma_3}{\mbox{min}}~\frac{1}{2}\sum_{k=t}^{t+T-1} \ell(u(k),\hat y(k),y_{r}(k)) + \Psi(\gamma_2,\gamma_3) \label{eq:cost_gammaDDPC}\\
			&~~\mbox{s.t.}~~\begin{bmatrix}
				u_{f}\\
				\hat y_{f}
			\end{bmatrix}=\begin{bmatrix}
				L_{21} & L_{22} & 0 \\
				L_{31} & L_{32} & L_{33}
			\end{bmatrix}\begin{bmatrix}
				\gamma_{1}^\star\\\gamma_{2} \\ \gamma_3
			\end{bmatrix} \label{eq:prediction_model3},\\
			&~~~~~~~~~u(k) \in \mathcal{U},~\hat y(k) \in \mathcal{Y},~k \in [t,t+T), \label{eq:constraints2}
		\end{align}
	\end{subequations}
	with 
        \begin{eqnarray}\label{eq:loss}
        \ell(u(k),\hat y(k),y_{r}(k))&=&\|\hat y(k)\!-\! y_{r}(k)
		\|_{Q}^{2}\!+\!\|u(k)\|_{R}^{2}, \\ \label{eq:init_terms}
  \gamma_{1}^\star&=&
			L_{11}^{-1}z_{init}
        \end{eqnarray}
where $z_{init}$ is defined as in \eqref{eq:zinit}. Note that, this formulation matches most\footnote{Some of the schemes in  \cite{dorfler2022bridging} require also a regularization on $
\gamma_1$, which instead in this work is  fixed based on \eqref{eq:init_terms}. See  \cite{breschi2022role} for further discussion about this issue.} of  the regularized DDPC schemes proposed in \cite{dorfler2022bridging}, each based on a specific choice of\footnote{Based on the relationship in \eqref{eq:LQ}, $\Psi(\gamma_2,\gamma_3)$ is indeed equivalent to a regularization on $\alpha$.} $\Psi(\gamma_2,\gamma_3)$. Nonetheless, by decoupling $\alpha$ in the three components $\gamma_1, \gamma_2$ and  $\gamma_3$, this formulation turns out to be more convenient  when discussing (and tuning) regularization.   


The regularization term $\Psi(\gamma_2,\gamma_3)$ in \MFa{\eqref{eq:cost_gammaDDPC}} is rather critical \SFo{to design}. Indeed, even in the presence of a small amount of noise, closed-loop performance might dramatically change depending on the chosen $\Psi(\gamma_2,\gamma_3)$. In the most extreme cases, one may go from not controlling the system at all (i.e., let it evolve in open loop) to overfitting noise, see e.g., \cite{dorfler2022bridging}. Based on these considerations, our ultimate {goal} is \textit{to provide a systematic approach for the design of the last term in \MFa{\eqref{eq:cost_gammaDDPC}} within our stochastic framework, while avoiding the need for additional experiments and closed-loop tuning tests}.

Note that some \VBr{observations} on the problem in \eqref{eq:RHPC_prob_dd_gamma} have already been \VBr{made} 
in \cite{breschi2022role}, where it is already argued that, for large  $N$, the optimal choice is to set $\gamma_3 =0$ and remove any regularization from $\gamma_2$, so that problem \eqref{eq:RHPC_prob_dd_gamma} becomes:	\begin{subequations}\label{eq:RHPC_prob_dd_gamma2_noreg}
		\begin{align}
			&\underset{\gamma_2}{\mbox{min}}~\frac{1}{2}\sum_{k=t}^{t+T-1} \!\!\!\|\hat y_{0}(k)\!-\! y_{r}(k)
		\|_{Q}^{2}\!+\!\|u(k)\|_{R}^{2}  \label{eq:cost}\\
			&~~\mbox{s.t.}~~\begin{bmatrix}
				u_{f}\\
				\hat y_{0,f}
			\end{bmatrix}=\begin{bmatrix}
				L_{21} & L_{22} \\
				L_{31} & L_{32} 
			\end{bmatrix}\begin{bmatrix}
				\gamma_{1}^\star\\\gamma_{2} 
			\end{bmatrix} \label{eq:prediction_model3},\\
			&~~~~~~~~~u(k) \in \mathcal{U},~\MFa{\hat{y}_{0}}(k) \in \mathcal{Y},~k \in [t,t+T), \label{eq:constraints2}
		\end{align}
	\end{subequations}
	Under this choice ($\gamma_3 = 0$) and with no regularization on $\gamma_2$, the predicted output $\hat y_{0,f} = L_{31} \gamma_1^\star + L_{32} \gamma_2$ can further be written (see \cite{breschi2022role}) in the form:
	\begin{equation}\label{eq:yhat}
	\begin{array}{rcl}
	\hat y_{0,f} & = & \begin{bmatrix}
				L_{31} & L_{32} 
			\end{bmatrix}\begin{bmatrix}
				\gamma_{1}^\star\\\gamma_{2} 
			\end{bmatrix}   = \hat Y_F \alpha
	\end{array}
	\end{equation}
	where the last equation exploits the fact that the projected future output $\hat Y_F$ 
	 can be written in terms of the LQ decomposition \eqref{eq:LQ} as:
	$$
	\hat Y_F =\begin{bmatrix}
				L_{31} & L_{32} 
			\end{bmatrix}\begin{bmatrix}
				Q_1\\ Q_2
			\end{bmatrix}.
	$$

\section{Finite sample uncertainty of data-driven predictors}\label{sec:uncertainty}

Exploiting the relation \eqref{eq:Proj:F}, 	the predicted output $\hat y_{0,f}$ in \eqref{eq:yhat} is subject to $O_p(1/\sqrt{N})$ perturbations that are due to the projection residuals
	of the future innovations $E_f$ onto the joint past $Z_p$ and future input $U_f$ spaces.

	More precisely, denoting with $\hat y^{*}_f$ the ``true'' output predictor corresponding to the given initial conditions and inputs, $\hat y_{0,f}$ satisfies the relation
	$$
	\underbrace{\hat y_{0,f}}_{ = \hat Y_f \alpha} = \underbrace{\hat y^{*}_f}_{= \left[\Gamma \hat X_\rho +{\cal H}_d U_F\right] \alpha} \!\!\!\!\!\!\!+ \MFa{\Hmc_s} \underbrace{\Pi_{Z_p,Uf}(E_f) \alpha}_{\tilde e_f},
	$$
	Defining the prediction error $\tilde y_f := \hat y^{*}_f - \hat y_{0,f} $, from the previous equation we see that
	$\tilde y_f  = -\MFa{\Hmc_s} \tilde e_f $.
If $\tilde y_f$ \MFa{were} equal to zero, then the optimal control problem  \eqref{eq:RHPC_prob_dd_gamma2_noreg} would coincide with the \emph{oracle} model based predictive control problem, i.e., the optimal MPC using the true model of the system \eqref{eq:RHPC_prob}. 

For future use, let us observe that $\tilde e_f$ can be written in the form: 
	$$
	\begin{array}{rcl}
	\tilde e_f&: =& \Pi_{Z_p,Uf}(E_f) \alpha =  E_f  \left[\begin{array}{ccc} Q^\top_1 &  Q^\top_2  \end{array}\right]   \left[\begin{array}{c} Q_1 \\ Q_2 \end{array}\right]  \alpha. 
\end{array}
	$$
	Denoting with $e_f(t)$ the $t-$th column of $E_f$ and with $q(t)$ the $t-$th column of
	 $$ \sqrt{N}\left[\begin{array}{c} Q_1 \\ Q_2 \end{array}\right]: = \left[ \begin{matrix} L_{11} & 0 \\ L_{21} & L_{22} \end{matrix}\right]^{-1}\sqrt{N}\left[\begin{array}{c} Z_P \\ U_F\end{array}\right],$$
	we can rewrite 	$\sqrt{N}\tilde e_f$ in the form: 
	$$
\sqrt{N}	\tilde e_f =\frac{1}{\sqrt{N}}\sum_{t=1}^{N} e_f(t) q(t)^\top  \underbrace{\left[\begin{array}{c} \gamma_1 \\ \gamma_2  \end{array}\right]}_{\ACh{:=\gamma_{12}}}.
        $$ 
        
The following proposition characterizes the statistical properties of $\tilde e_f$, and is the core result that will be used in the next section to design data-driven tuning strategies for regularization in DDPC.

\begin{prop}\label{Prop:eftilde}
Under the assumption that the innovation process $e(t)$ in \eqref{eq:stoc_sys} is, conditionally on the joint input-output past data $\{y(s),u(s), s <t\}$,  a martingale difference sequence with constant conditional variance, i.e.,
$$
\begin{array}{c}
\E[e(t) |y(s),u(s), s <t] = 0,\\
Var\MFa{[ }e(t) |y(s),u(s), s <t\MFa{]} =Var\MFa{[ }e(t)\MFa{]} =\sigma^2,
\end{array}
$$
then 
$$
\E \VBr{[}\sqrt{N} \tilde e_f\VBr{]}\mathop{\longrightarrow}^{N\rightarrow \infty}0,
$$
and 
\begin{equation}\label{eq:Varetilde}
Var\MFa{[ } \sqrt{N}\tilde e_f\MFa{]} \mathop{\longrightarrow}^{N\rightarrow \infty} 
\sum_{k=-T}^{T}\sigma^2 J_{k} \frac{(N-|k|)}{N} \ACh{\gamma_{12}}^\top  \Sigma^\top_q(k) \ACh{\gamma_{12}},
\end{equation}
 where $\Sigma_q(k)$ is the covariance matrix  $\E \MFa{[ } q(t+k) q^\top(t)\MFa{] }$ and 
 $J_i$ is the  shift  matrix   such that $[J_i]_{h,k} = 0$ for $k-h \neq i$ 
 and $[J_i]_{h,k} =1$ for $k-h = i$ (i.e., with zeros everywhere except for ones on the superdiagonal ($i>0$) or subdiagonal ($i<0$)).
\end{prop}

The covariance matrices $\Sigma_q(k)$ in \eqref{eq:Varetilde} can be estimated from data. However, it is easy to prove that, asymptotically in $N$, $\Sigma_q(0) = I$. Therefore, exploiting the fact that $Trace\MFa{[ }J_i\MFa{] } = 0$ for $i\neq 0$, we have that 
\begin{equation}\label{eq:ScalarVarianceetilde}
Trace\MFa{[ }Var\MFa{[ } \sqrt{N}\tilde e_f\MFa{] }\MFa{] } \mathop{\longrightarrow}^{N\rightarrow \infty} 
T \sigma^2 \|\ACh{\gamma_{12}}\|^2.
\end{equation}

This relation will be extremely useful in the next section because it connects  the average scalar variance of $\tilde e_f$ to the optimization variable $\gamma$. In particular, if the control  horizon $T$  is large enough, one expects that $\frac{\|\tilde e_f\|^2}{T}$ can be seen as a sample estimate of the average variance $\frac{Trace\MFa{[ }Var\MFa{[ } \tilde e_f\MFa{] }\MFa{] }}{T}$ of the components of the vector $\tilde e_f$, so that:
$$
\frac{\|\tilde e_f\|^2}{T} \simeq \frac{Trace\MFa{[ }Var\MFa{[ } \tilde e_f\MFa{] }\MFa{] }}{T} \simeq \frac{\sigma^2 \|\ACh{\gamma_{12}}\|^2}{N}. 
$$

We 
conclude this section discussing how the connection between the prediction error $\tilde y_f$ and the vector $\tilde e_f$, given by $\tilde y_f = -\MFa{\Hmc_s} \tilde e_f$ can be estimated in a model-free, data driven fashion. The result is formalized in the following lemma.

\begin{lem}
Consider the LQ decomposition in \eqref{eq:LQ}. Then 
\begin{equation}\label{eq:ToeplitzVariance}
\mathop{\rm lim}_{N\rightarrow \infty} L_{33}L_{33}^\top = \sigma^2 \MFa{\Hmc_s} \MFa{\Hmc_s}^\top.
\end{equation}
\end{lem}

The previous lemma shows that the matrix $L_{33}$ can be seen as a (nonparametric/model-free) estimate of $\sigma \MFa{\Hmc_s}$. Exploiting this fact we obtain the fundamental result of this section that will be used later on \VBr{in the design of our tuning strategy.}
\begin{prop}\label{prop:ytilde}
The prediction error $\tilde y_f$ can be written as 
$$
\tilde y_f = -\MFa{\Hmc_s} \tilde e_f \simeq - L_{33} \frac{\tilde e_f }{\sigma} = L_{33} \tilde \gamma_3,
$$
where $\tilde \gamma_3:= -\frac{\tilde e_f }{\sigma} $ satisfies
$$
\|\tilde \gamma_3 \|^2 \simeq Trace\MFa{[ }Var\MFa{[ } \tilde  \gamma_3\MFa{] }\MFa{] }\simeq \frac{T\|\ACh{\gamma_{12}}\|^2}{N}. 
$$
\end{prop}

%

\section{Regularization design}\label{sec:reg}

In Section \ref{sec:uncertainty} , we have seen that (i) the predictor 
$\hat y_f$ used in the data-driven predictive control design  scheme \eqref{eq:RHPC_prob_dd_gamma2_noreg} is affected by an uncertainty $\tilde y_f$, (ii) how the latter can be (statistically) characterized, and (iii)  how it can be expressed in a data-driven (model-free) fashion. Based on these considerations, in the next two subsections, we 
introduce two regularization strategies to mitigate the effect of the prediction error $\tilde y_f$ on the control performance.

\subsection{Regularizing $\gamma_2$}\label{ssec:reg_gamma2} First of all let us observe that the average scalar variance \VBr{$Trace[Var[\gamma_3]]$}
of the vector $\gamma_3: = L^{-1}_{33} \tilde y_f$ scales linearly with squared norm of the optimization parameter $\|\ACh{\gamma_{12}}\|^2 = \|\gamma_1\|^2 + \|\gamma_2\|^2$. The vector $\gamma_1$ is  fixed in the optimization problem \eqref{eq:RHPC_prob_dd_gamma} to \VBr{the }
value $\gamma_1^*$ that  matche\VBr{s} 
the initial conditions, whereas $\gamma_2$ is optimized to achieve the control goal. Should $\|\gamma_2\|$ grow, then also the variance of the prediction error would increase, possibly jeopardizing the closed loop performance. Thus, it is  desirable to regularize the control problem \eqref{eq:RHPC_prob_dd_gamma} by adding a term of the form $\Psi(\gamma_2,\gamma_3) = \beta_2 \|\gamma_2\|^2$ while  constraining $\gamma_3 =0$, i.e.,
\begin{subequations}\label{eq:RHPC_prob_dd_gamma2}
		\begin{align}
			&\underset{\gamma_2}{\mbox{min}}~\frac{1}{2}\sum_{k=t}^{t+T-1} \ell(u(k),\MFa{\hat{y}_{0}}(k),y_{r}(k)) + \beta_2 \|\gamma_2\|^2 \label{eq:cost}\\
			&~~\mbox{s.t.}~~\begin{bmatrix}
				u_{f}\\
				\MFa{\hat{y}}_{0,f}
			\end{bmatrix}=\begin{bmatrix}
				L_{21} & L_{22} \\
				L_{31} & L_{32} 
			\end{bmatrix}\begin{bmatrix}
				\gamma_{1}^\star\\\gamma_{2} 
			\end{bmatrix} \label{eq:prediction_model3},\\
	   &~~~~~~~~~u(k) \in \mathcal{U},~\MFa{\hat{y}_{0}}(k) \in \mathcal{Y},~k \in [t,t+T), \label{eq:constraints2}
		\end{align}
	\end{subequations}
    where $\ell(u(k),\MFa{\hat{y}_{0}}(k),y_{r}(k))$ is defined as in \eqref{eq:loss}. This scheme is designed so as to keep the norm of $\gamma_2$ (and thus the variance of the prediction error) small. In particular, we would like to avoid situations in which the control input is (erroneously) exploiting prediction errors to make $\hat y_{0,f}$ (too) close to the reference trajectory $y_r(t)$ (\VBr{indeed, there are }no reasons  to fit below noise level).  Thus, it makes sense to chose $\beta_2$ large enough so that 
    \begin{equation}\label{eq:est:beta2}
    \|L_{33}^{-1}(\hat y_{0,f} - y_r)\|^2 \simeq \|L_{33}^{-1}\tilde y_f\|^2  \simeq \MFa{T} \frac{\|\gamma_1^*\|^2 + \|\gamma^*_2(\beta_2)\|^2}{N},\end{equation}
    were $\gamma^*_2(\beta_2)$ denotes the optimal parameter $\gamma_2$ that solve \eqref{eq:RHPC_prob_dd_gamma2} as a function of $\beta_2$.
    As \VBr{it will be shown} 
    in Section~\ref{sec:examples}, this reduces to a linear search problem that can be solved \emph{before} actually closing the loop. 


\subsection{Slack on output prediction}\label{ssec:slack_gamma3} 

As an alternative,  Proposition \ref{prop:ytilde} suggests that the ``true'' predictor $\hat y_f^{*}$ can be written as
$$
\hat y_f^{*} = \hat y_{0,f} + L_{33} \tilde \gamma_3,
$$
where $\tilde  \gamma_{\MFa{3}}$ depends on the ``true'' but unknown noise. This suggests that the effect of noise can be compensated for by adding a slack of the form  $  \xi_f: =  L_{33}  \gamma_3$ with $\gamma_3$ to be optimized.
We do so by defining 
\begin{equation*}
	\hat y_{f} := \hat y_{0,f} + \xi_{f} = L_{11} \gamma_{1}^{\star} + L_{22} \gamma_{2} + L_{33} \gamma_{3},
\end{equation*}
where $\gamma_{2}$ and $\gamma_{3}$ are optimization variables.
In this case, introducing some slack has the effect of avoiding an unnecessarily  large control effort, thus inducing an implicit regularization on $\gamma_2$.  The ``size'' of the slack variable can be controlled by regularizing the norm of $\gamma_3$, adding a regularization term $\Psi(\beta_2,\beta_3):=\beta_3 \|\gamma_3\|^2$ \VBr{in \eqref{eq:RHPC_prob_dd_gamma}}. The resulting optimization problem \ACh{(changing sign to the optimization variable $\gamma_3$, which of course does not affect the result)} takes the form 
		\begin{subequations}\label{eq:RHPC_prob_dd_gamma3}
		\begin{align}
			&\underset{\gamma_2,\gamma_3}{\mbox{min}}~\frac{1}{2}\sum_{k=t}^{t+T-1} \ell(u(k),\MFa{\hat{y}}(k),y_{r}(k)) + \beta_3 \|\gamma_3\|^2 \label{eq:cost}\\
			&~~\mbox{s.t.}~~\begin{bmatrix}
				u_{f}\\
				\MFa{\hat{y}}_{f}
			\end{bmatrix}=\begin{bmatrix}
				L_{21} & L_{22} & 0 \\
				L_{31} & L_{32} & L_{33}
			\end{bmatrix}\begin{bmatrix}
				\gamma_{1}^\star\\\gamma_{2} \\ \gamma_3
			\end{bmatrix} \label{eq:prediction_model3},\\
			&~~~~~~~~~u(k) \in \mathcal{U},~\MFa{\hat{y}}(k) \in \mathcal{Y},~k \in [t,t+T), \label{eq:constraints2}
		\end{align}
	\end{subequations}
 with $\ell(u(k),\MFa{\hat{y}}(k),y_{r}(k))$ given by \eqref{eq:loss}.  Also in this case $\beta_3$ can be tuned, via a linear search, to be small enough\footnote{Note that here $\beta_3 \rightarrow \infty$ implies that $\gamma_3 = 0$ and thus no slack would be introduced.} so as to guarantee that
\begin{equation}\label{eq:est:beta3}
   \|\gamma^\star_3(\beta_3)\|^2 \simeq \MFa{T} \frac{\|\gamma_1^\star\|^2 + \|\gamma^\star_2(\beta_3)\|^2}{N},\end{equation}
    were $\gamma^\star_2(\beta_3)$ and $\gamma^\star_3(\beta_3)$  denote the optimal parameter $\gamma_2$ and $\gamma_3$ that solve \eqref{eq:RHPC_prob_dd_gamma3} as a function of $\beta_3$.


\begin{figure*}[t!] 
	\centering
    \subfigure[]{\includegraphics[width=0.234\textwidth,trim={0.1cm 0.3cm 0.8cm 0.4cm},clip]{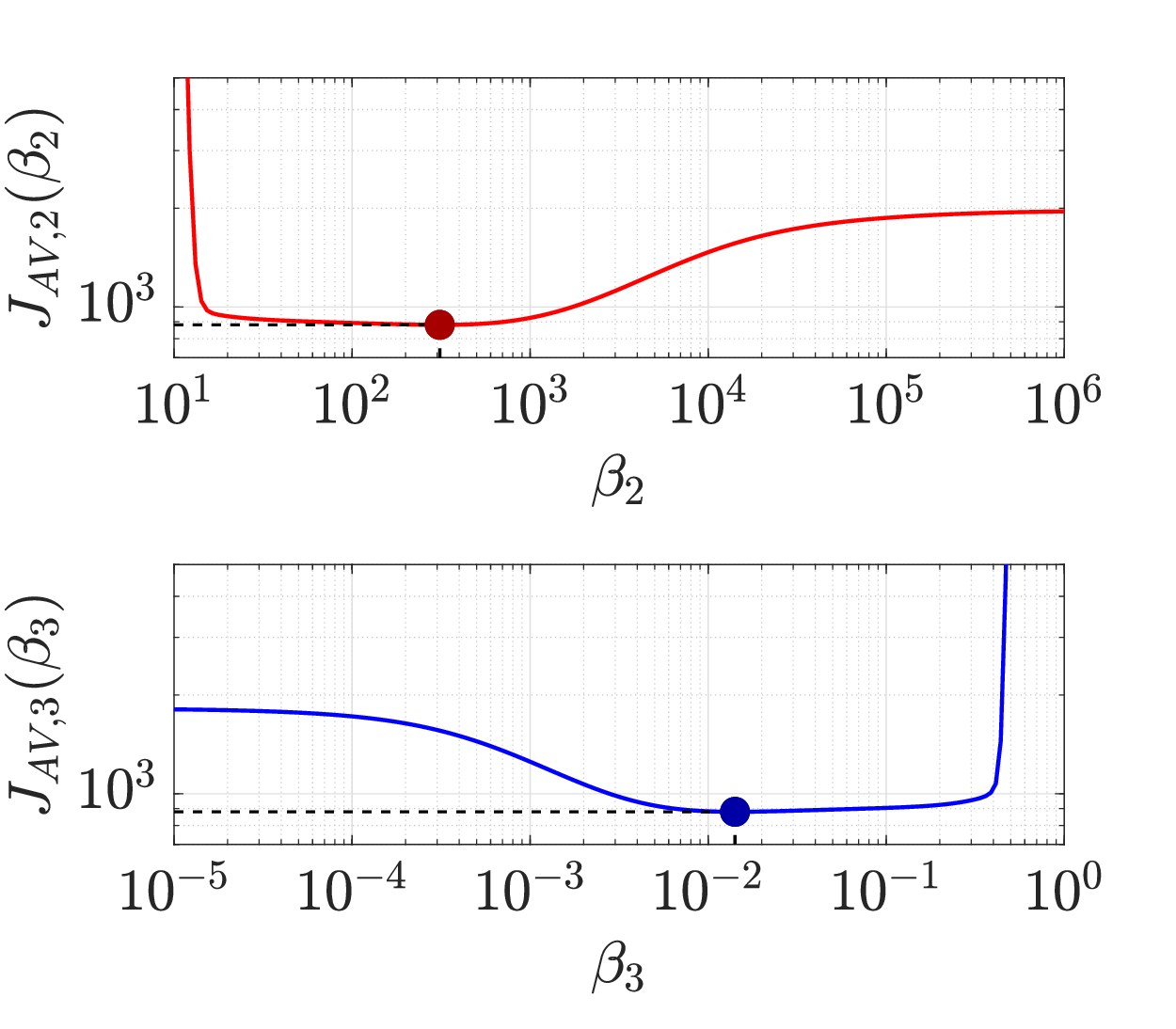}\label{fig:Jave_}}
	\subfigure[]{\includegraphics[width=0.244\textwidth,trim={0.3cm -0.2cm 2cm 0},clip]{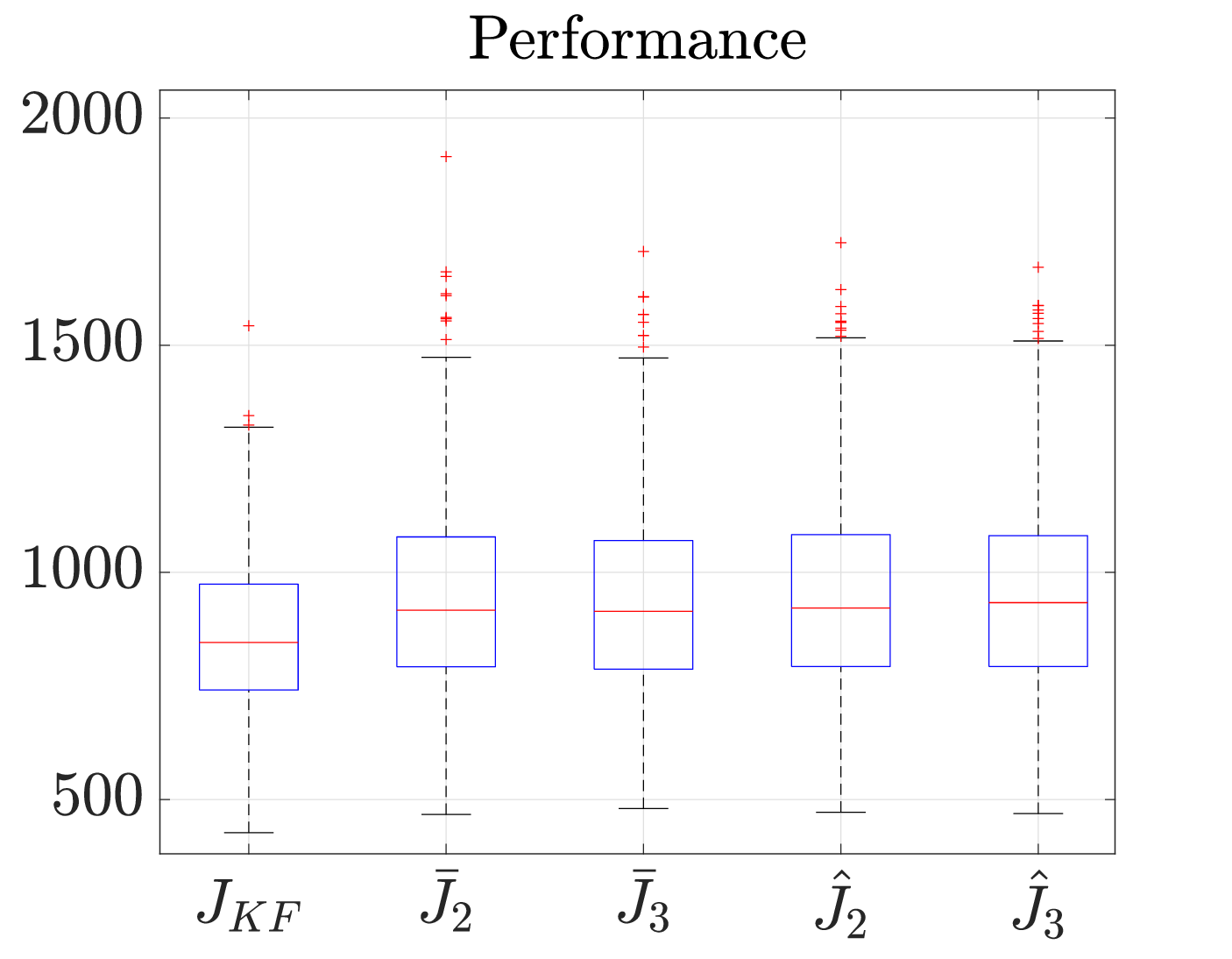}\label{fig:allJ}} 
	\subfigure[]{\includegraphics[width=0.244\textwidth,trim={0.3cm -0.2cm 2cm 0},clip]{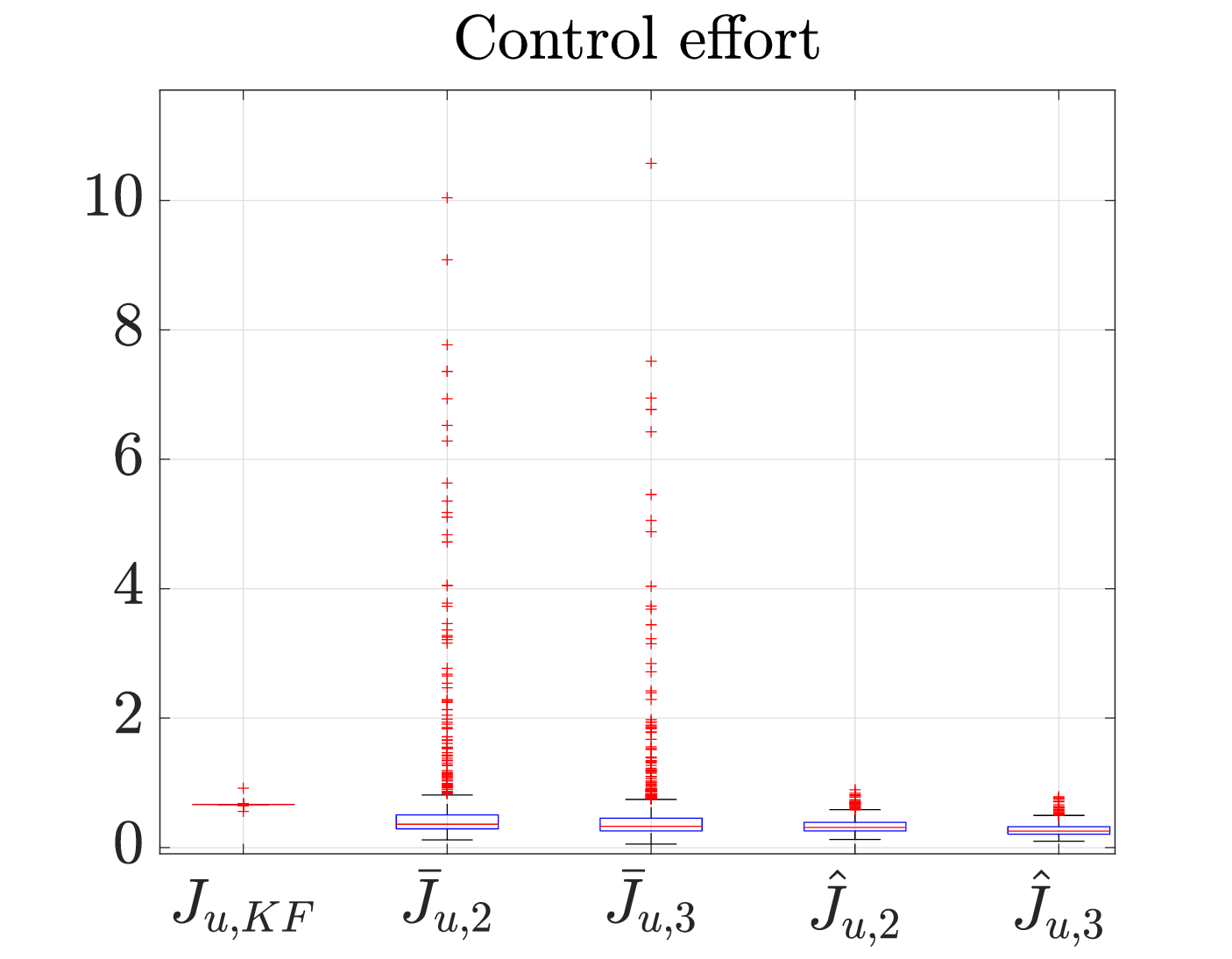}\label{fig:allJu}} 
	\subfigure[]{\includegraphics[width=0.244\textwidth,trim={0.3cm -0.2cm 2cm 0},clip]{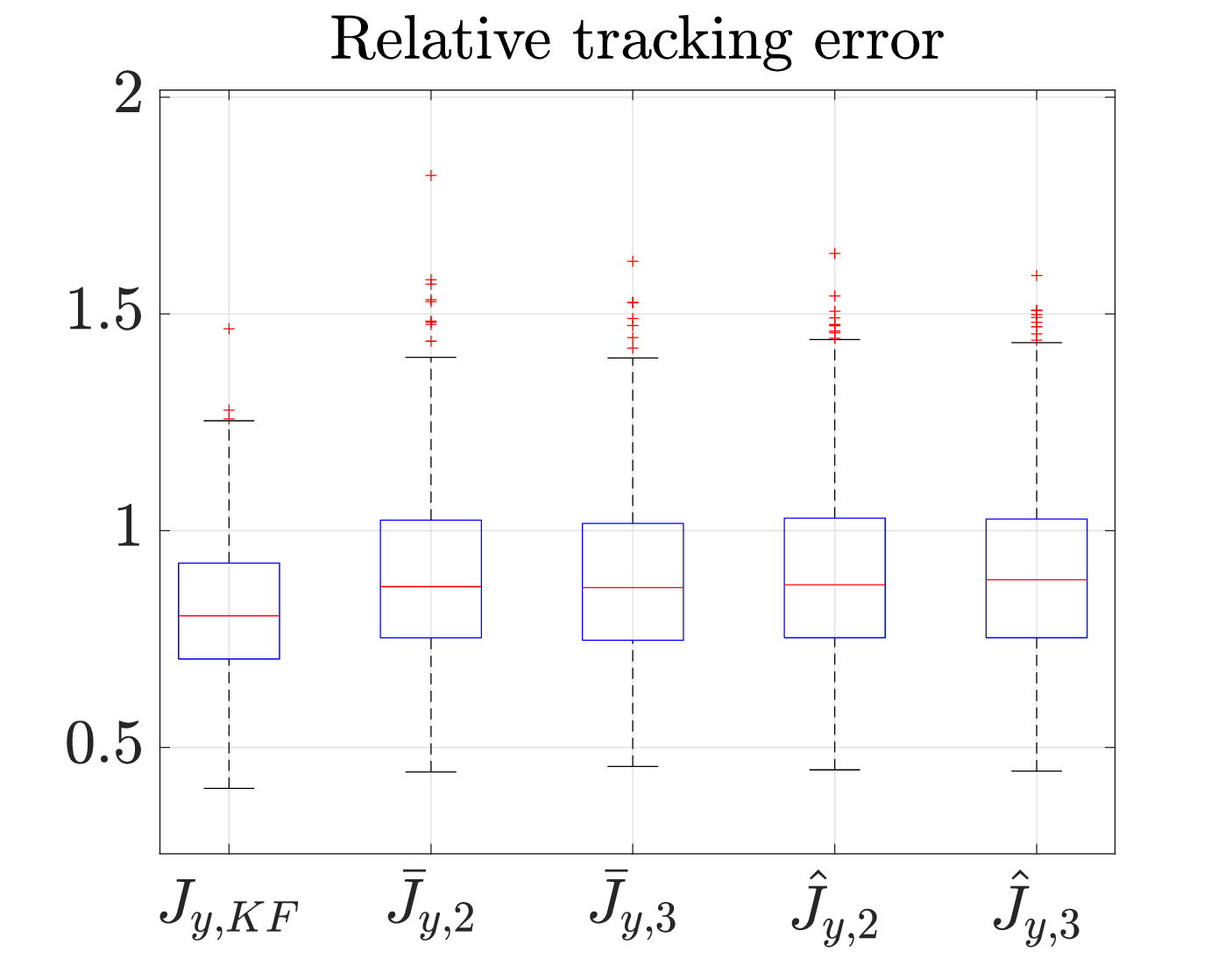}\label{fig:allJy}}
	\vspace{-.2cm}
	\caption{(a): Oracle tuning: average  closed-loop costs $J_{AV,a}$ and their  minimizers $\bar{\beta}_a$, $a=2,3$.  (b),(c),(d): Distributions of the closed loop performance, control effort and relative tracking error over  $1000$ Monte Carlo runs. Costs with bars refer to ``oracle'' tuning  whereas costs with hats refer to online (feasible) strategy.}
	\label{fig:allJs}
\end{figure*}


\section{Numerical examples}\label{sec:examples}

We \VBr{now} 
report 
\VBr{the results of }some numerical \VBr{simulations} 
to illustrate our theoretical findings, by considering the benchmark single-input, single-output, $4$-th order, linear time-invariant system in \cite{LandauReyKarimi1995}. 
Similarly to  \cite{dorfler2022bridging}, we collect one noise-free input/output time series of length $N_{data}=250$, by applying a random Gaussian input of unitary variance. From this noise-free data set, $n_{MC}=1000$ independent noisy data sets are constructed by adding Gaussian noise with signal-to-noise ratio of $13$ dB, to be used as training data for the closed loop experiments. The latter  are carried out in a \emph{noisy} scenario, i.e., wit noise  added to the output and thus fed back in the loop.


In the experiments the output reference\footnote{The data-driven predictive controller is designed with preview of the reference to be tracked.} is $y_r(t)=\sin(5 \pi t / (T+T_v-1))$ the prediction horizon $T=20$ and  the number of feedback steps is $T_v = 50$. The performance index we use is chosen as
\begin{equation}\label{eq:perfindex}
	J_a = T_{v}^{-1} \sum\nolimits_{t=0}^{T_{v}-1} \left\|y(t)-y_{r}(t)\right\|_{Q}^{2}+\left\|u(t)\right\|_{R}^{2},
\end{equation}
where $(Q,R) = (qI_{p},rI_{m})$, with $(q,r) = (2000,0.01)$ and $m=p=1$. The subscript $a=2$ refers to schemes regularizing $\gamma_2$ as discussed in Section  \ref{ssec:reg_gamma2}, while $a=3$ is associated with approaches tuning $\gamma_3$ 
(see Section \ref{ssec:slack_gamma3}). 
To benchmark 
closed-loop performance, we consider a Kalman-filter-based model predictive scheme that exploits the true model parameters, and denote its performance index as $J_{a}$, $a=KF$.
%
We also report the corresponding  control effort $$J_{u,a} = T_{v}^{-1} \sum\nolimits_{t=0}^{T_{v}-1} \left\|u(t)\right\|^2,$$ and   relative tracking error $$J_{y,a} = \left(\sum\nolimits_{t=0}^{T_{v}-1}\left\|y(t)-y_{r}(t)\right\|^2\right) / \left( \sum\nolimits_{t=0}^{T_{v}-1} \left\|y_r(t)\right\|^2\right),$$ where $a$ is either $2$, $3$, or $KF$. 

%


We benchmark our tuning strategies for $\beta_2$ and $\beta_3$ against an oracle that exploits an estimate of the close-loop cost obtained as follows by running  $1000$ closed-loop Monte Carlo trials and then averaging: $J_{AV,a} = n_{MC}^{-1} \sum_{j=1}^{n_{MC}} J^{(j)}_a$.  $J^{(j)}_a$ denotes  
the $j$-th realization of the cost $J_a$ and it is computed for 
 $\beta_a$  ($a=2,3)$ ranging on a \MFa{logarithmically} spaced grid $G_a$ of $|G_a|=200$ fixed points for $\beta_2 \in G_2 \subset [10^{0},10^{4}]$ and $\beta_3 \in G_3 \subset [10^{-4},10^{0}]$. The corresponding minimizers are denoted as  $\bar{\beta}_a$, $a\in\{2,3\}$ (see \figurename{~\ref{fig:Jave_}}) \footnote{Note that the values resulting from this procedure cannot be retrieved in practice without performing a set of (possibly unsafe) closed loop experiments.}.


Closed loop performance are evaluated on $n_{MC}$ Monte Carlo 
experiments with the following four configurations. 
\begin{enumerate}
    \item[(a)]  The regularized problem in Section \ref{ssec:reg_gamma2} with  $\beta_2(t) = \bar{\beta}_2$. The corresponding costs are denoted as $\bar J_{2}$.
    \item[(b)] The regularized problem in Section \ref{ssec:slack_gamma3}  with   $\beta_3(t) = \bar{\beta}_3$. The corresponding costs are denoted as $\bar J_{3}$.
    \item[(c)]   The regularized problem in Section \ref{ssec:reg_gamma2} is solved with $\beta_2(t)$ tuned at each closed loop step  enforcing condition \eqref{eq:est:beta2}. The corresponding costs are denoted as $\hat J_{2}$.
    \item[(d)]   The regularized problem in Section \ref{ssec:slack_gamma3} with $\beta_3(t)$ tuned at each close loop step  enforcing condition \eqref{eq:est:beta3}. The corresponding costs are denoted as $\hat J_{3}$.
    \end{enumerate}

In the sequel, we refer to the selection $\beta_a(t) = \bar{\beta}_a$ in (a) and (b) as the ``offline'' tuning strategies; whereas, (c) and (d) are the ``online'' tuning approaches (as they do not require additional closed-loop experiments)  proposed in this paper. 

Figg. \ref{fig:allJ}-\ref{fig:allJy} show the boxplots  $n_{MC}$ Monte Carlo runs of the realized closed loop costs. 
Remarkably, the offline selection of $\beta_a(t) = \bar{\beta}_a$ and the online strategies using  \eqref{eq:est:beta2} and \eqref{eq:est:beta3} perform comparably. \\ 
Furthermore,  \figurename{~\ref{fig:Jave_}} confirms that, on average,   the regularizing role of the parameters $\beta_2$ and $\beta_3$ is crucial \VBr{for the} 
stability (and optimal performance) of the closed loop. It is worth observing that $\beta_2$ and $\beta_3$ play dual roles. Indeed, $\beta_2 \rightarrow 0$ and $\beta_3 \rightarrow +\infty$ correspond to no regularization on the control problem, whereas  $\beta_2 \rightarrow \MFa{+}\infty$ and $\beta_3 \rightarrow 0$ correspond to ``maximal'' regularization. Lastly, from \figurename{~\ref{fig:Jave_}} it is clear that  the closed-loop control cost diverges (a clear sign of instability), when no regularization is performed, i.e.,  $\beta_2 \rightarrow 0$ or $\beta_3 \rightarrow +\infty$.

\section{Conclusions}\label{sec:concl}

Leveraging the statistical analysis of the (non-parametric) predictor used in 
data-driven, model-free, predictive control problems, in this paper we have proposed two regularization approaches to account for finite sample effects in the design of data-driven predictive controllers within a stochastic setting. We have also discussed  corresponding online tuning strategies for the selection of the regularization penalties. The proposed tuning rationale allows for the design of the controller without the need for additional closed-loop experiments. 

Simulation results confirm the effectiveness of the online strategies in 
face of uncertainties, showing 
that their performance is \SFo{practically} indistinguishable from an oracle-type tuning based on offline \SFo{closed-loop} experiments. Future work will include a thorough evaluation of the proposed on-line tuning strategy, as well as a formal analysis of the closed loop stability.


\bibliography{biblio}

\begin{thebibliography}{13}
\providecommand{\natexlab}[1]{#1}
\providecommand{\url}[1]{\texttt{#1}}
\providecommand{\urlprefix}{URL }
\expandafter\ifx\csname urlstyle\endcsname\relax
  \providecommand{\doi}[1]{doi:\discretionary{}{}{}#1}\else
  \providecommand{\doi}{doi:\discretionary{}{}{}\begingroup
  \urlstyle{rm}\Url}\fi

\bibitem[{Berberich et~al.(2020)Berberich, K{\"o}hler, M{\"u}ller, and
  Allg{\"o}wer}]{berberich2020data}
Berberich, J., K{\"o}hler, J., M{\"u}ller, M.A., and Allg{\"o}wer, F. (2020).
\newblock Data-driven model predictive control with stability and robustness
  guarantees.
\newblock \emph{IEEE Transactions on Automatic Control}, 66(4), 1702--1717.

\bibitem[{Borrelli et~al.(2017)Borrelli, Bemporad, and
  Morari}]{borrelli2017predictive}
Borrelli, F., Bemporad, A., and Morari, M. (2017).
\newblock \emph{Predictive control for linear and hybrid systems}.
\newblock Cambridge University Press.

\bibitem[{Breschi et~al.(2023)Breschi, Chiuso, and Formentin}]{breschi2022role}
Breschi, V., Chiuso, A., and Formentin, S. (2023).
\newblock Data-driven predictive control in a stochastic setting: a unified
  framework.
\newblock \emph{Automatica}, 152(110961).

\bibitem[{Coulson et~al.(2019)Coulson, Lygeros, and
  D{\"o}rfler}]{coulson2019data}
Coulson, J., Lygeros, J., and D{\"o}rfler, F. (2019).
\newblock Data-enabled predictive control: In the shallows of the deepc.
\newblock In \emph{2019 18th European Control Conference (ECC)}, 307--312.
  IEEE.

\bibitem[{Ding(2017)}]{ding2017robust}
Ding, B. (2017).
\newblock Robust and adaptive model predictive control of nonlinear systems.
\newblock \emph{IEEE Control Systems Magazine}, 37(1), 125--127.

\bibitem[{Dorfler et~al.(2022)Dorfler, Coulson, and
  Markovsky}]{dorfler2022bridging}
Dorfler, F., Coulson, J., and Markovsky, I. (2022).
\newblock Bridging direct \& indirect data-driven control formulations via
  regularizations and relaxations.
\newblock \emph{IEEE Transactions on Automatic Control}.

\bibitem[{Favoreel et~al.(1999)Favoreel, Moor, and Gevers}]{FAVOREEL1999}
Favoreel, W., Moor, B.D., and Gevers, M. (1999).
\newblock Spc: Subspace predictive control.
\newblock \emph{IFAC Proceedings Volumes}, 32(2), 4004--4009.
\newblock 14th IFAC World Congress 1999, Beijing, Chia, 5-9 July.

\bibitem[{Fiedler and Lucia(2021)}]{Felix2021}
Fiedler, F. and Lucia, S. (2021).
\newblock On the relationship between data-enabled predictive control and
  subspace predictive control.
\newblock In \emph{2021 European Control Conference (ECC)}, 222--229.

\bibitem[{Krishnan and Pasqualetti(2021)}]{krishnan2021direct}
Krishnan, V. and Pasqualetti, F. (2021).
\newblock On direct vs indirect data-driven predictive control.
\newblock In \emph{2021 60th IEEE Conference on Decision and Control (CDC)},
  736--741. IEEE.

\bibitem[{Landau et~al.(1995)Landau, Rey, Karimi, Voda, and
  Franco}]{LandauReyKarimi1995}
Landau, I., Rey, D., Karimi, A., Voda, A., and Franco, A. (1995).
\newblock A flexible transmission system as a benchmark for robust digital
  control.
\newblock \emph{European Journal of Control}, 1(2), 77--96.

\bibitem[{Moonen et~al.(1989)Moonen, {De~Moor}, Vandeberghe, and
  Vandewalle}]{Moonen}
Moonen, M., {De~Moor}, B., Vandeberghe, L., and Vandewalle, J. (1989).
\newblock On- and off-line identification of linear state-space models.
\newblock \emph{Int. J. of Control}, 49(1), 219--232.

\bibitem[{Willems et~al.(2005)Willems, Rapisarda, Markovsky, and
  De~Moor}]{willems2005note}
Willems, J.C., Rapisarda, P., Markovsky, I., and De~Moor, B.L. (2005).
\newblock A note on persistency of excitation.
\newblock \emph{Systems \& Control Letters}, 54(4), 325--329.

\bibitem[{Willems and Polderman(2013)}]{willems2013introduction}
Willems, J. and Polderman, J. (2013).
\newblock \emph{Introduction to Mathematical Systems Theory: A Behavioral
  Approach}.
\newblock Texts in Applied Mathematics. Springer New York.
\newblock \urlprefix\url{https://books.google.it/books?id=qoLSBwAAQBAJ}.

\end{thebibliography}


\end{document}